\documentclass[quantumrep,article,accept,pdftex,oneauthor]{Definitions/mdpi} 

\firstpage{1} 
\pubvolume{1}
\issuenum{1}
\articlenumber{0}
\pubyear{2024}
\copyrightyear{2024}
\externaleditor{Academic Editors: Viktor Dodonov, Margarita A. Man’ko, Salomon S Mizrahi, Luis L. Sánchez-Soto}
\datereceived{10 July 2024} 
\daterevised{14 August 2024} 
\dateaccepted{19 August 2024} 
\datepublished{ } 
\hreflink{https://doi.org/} 

\usepackage[english]{babel}
\usepackage[mathscr]{euscript}
\usepackage{amsmath}
\usepackage{amsfonts,amssymb}
\usepackage{graphicx}

\usepackage[font=footnotesize,labelfont=]{caption}
	\newcommand{\bg}{\begin{linenomath*}\begin{eqnarray}}
			\newcommand{\ed}{\end{eqnarray}\end{linenomath*}}

\Title{Spin Helicity and the Disproof of Bell's Theorem}

\TitleCitation{Spin Helicity and the Disproof of Bell's Theorem}

\Author{Bryan Sanctuary 
 $^\dagger$}

\AuthorNames{Bryan Sanctuary}

\AuthorCitation{{Sanctuary, B.}} 

\address[1]{Chemistry Department, McGill University, Montreal, QC H3A 0G4, Canada; bryan.sanctuary@mcgill.ca 
}
\firstnote{\hangafter=1 \hangindent=1.05em \hspace{-0.82em}Retired  Professor.} 
	
\abstract{Under the quaternion group, $Q_8$,  spin helicity emerges as a crucial element of the reality  of spin and is complementary to its polarization.  We show that the correlation in EPR coincidence experiments is conserved upon separation from a singlet state and  distributed between its polarization and coherence. Including helicity accounts for the violation of Bell's Inequalities without non-locality, and disproves Bell's Theorem by a counterexample.}
		
\keyword{helicity; quantum foundations; Bell theorem; Bell's inequalities; spin theory; entanglement; complexification; geometric algebra; non-locality; spin coherence} 		

\begin{document}	
\section{Introduction}

Spin helicity, as defined in particle physics \cite{griffin}, is the projection of spin along the axis of linear momentum, $\mathbf{p}\cdot \sigma $. If the helicity is positive, the spin state is $\left| + \right\rangle $, and if negative, the state is $\left| - \right\rangle$.  In this paper, we abandon this definition. 

A significant difference between quantum and classical systems is that the former have complementary properties, like position and momentum, ${{\left[ p,r \right]}_{-}}=-i\hbar $, where $p=-i\hbar \frac{d}{dr}$, whereas classical systems all commute. The complementary property of spin polarization has not been formulated (see,  however, \cite{sanctuary1}), and here we find it is the spin helicity.  By complementary, we mean that distinct inverse spaces exist for the two properties in addition to not commuting. Position--momentum are represented by inverse Fourier spaces. That spin components do not commute, ${{\left[ {{\sigma }_{i}},{{\sigma }_{j}} \right]}_{-}}=2{{\varepsilon }_{ijk}}i{{\sigma }_{k}}$, does not make $\sigma_i$ and $\sigma_j$ complementary because they belong to the same vector space. Rather, these two components are incompatible. Helicity is the complementary attribute to spin polarization in a different complementary space.

This paper is the second of four which discusses changing the symmetry of spin from SU(2) to the quaternion group, $Q_8$.  We call the resulting spin Q-spin.  In the first paper, \cite{sanctuary1}, we give the formal development by recognizing that the Dirac equation, \cite{dirac}, does not include a bivector.  We introduce one by multiplying a gamma matrix by the imaginary number $i$. This complexifies spin spacetime in a similar manner to Twistor theory, \cite{penrose1,penrose2}, and leads to the definition of spin helicity. One consequence is that Q-spin has structure and is not a point particle. Rather than the matter--antimatter pair Dirac proposed, \cite{dirac2}, that is two particles with two states each, Q-spin has structure, being one particle with four~states. 

Here, we analyze the geometry of an EPR pair after separation from a singlet state. This shows that the off-diagonal elements of the density matrix are responsible for the quantum coherence that leads to the violation of Bell's Inequalities, (BI), \cite{clauser,aspect,aspect2,zeilinger},  without the need for non-local connections between Alice and Bob. This disproves Bell's Theorem,~\cite{BellBert}, by a counterexample. Additionally, we show that upon separation from a singlet, which is devoid of entanglement, correlation is conserved, being divided between polarization and coherence. We call this the Conservation of Geometric Correlation.

However, because of the importance and far-reaching consequences of Bell's Theorem, more evidence is needed to confirm its demise. In the third paper, \cite{sanctuary3}, the disproof is confirmed by a computer simulation that reproduces the observed violation of Bell's Inequalities. Simulations give deeper insight into the processes, \cite{Werner, Cabello,Summers,Leggett,Valentini,hess}. From our classical simulation, we find that the correlation has a CHSH, \cite{clauser}, a value of three, and not $2\sqrt{2}=2.828$. We discuss this and other features, such as the violation of the Tsirelson's bound, \cite{Tsirel}, and show that Q-spin gives a mechanism for quantum weirdness, \cite{mullin}.  The fourth paper, \cite{sanctuary4}, explores other consequences of Q-spin on the foundations of physics.

\subsection{The Singlet State and Separation} \label{sec9}

Our view of a singlet state is based on the cancellation of polarization. Looking at the usual definition of the singlet, we express two generic states each for Alice and Bob, ${{\vert \pm  \rangle }_{i}}$, by the up/down polarized states, and so the singlet state is constructed in the usual way:
\begin{equation} \label{E4}
	{{\vert \Psi_{12}  \rangle }}=\frac{1}{\sqrt{2}}\left[ {{\vert + \rangle }_{1}}{{\vert - \rangle }_{2}}-{{\vert - \rangle }_{1}}{{\vert + \rangle }_{2}} \right]
\end{equation}	
The states for each spin display orthogonalities: $\langle m \vert n\rangle=\delta_{mn}$. By generic, we mean the spin orientation is not specified, and any direction may emerge as long as it is the same for both Alice and Bob.

State vectors, like  Equation (\ref{E4}), do not explicitly display coherence, so taking its outer product gives a $4\times4$ matrix, which is the state operator, \cite{fano}:
\begin{equation}\label{G1}
	\begin{aligned}
		{{\rho }_{12}}
		& =\left| {{\Psi }_{12}} \right\rangle \left\langle  {{\Psi }_{12}} \right|\\
		& =\frac{1}{4}\left( {{I}^{1}}{{I}^{2}}-{{\sigma }^{1}}\cdot {{\sigma}^{2}} \right) 
		=\frac{1}{2}\left( \begin{matrix}
			0 & 0 & 0 & 0  \\
			0 & 1 & -1 & 0  \\
			0 & -1 & 1 & 0  \\
			0 & 0 & 0 & 0  \\
		\end{matrix} \right) \\ 
	\end{aligned}	    
\end{equation}
The polarized states are diagonal, $\left| \pm  \right\rangle \left\langle  \pm  \right|$, and the coherent states are off-diagonal, $\left| \pm  \right\rangle \left\langle  \mp  \right|$.   The resulting matrix is the entangled singlet state, $\rho_{12}$, and can be represented as the tensor product between the two identity matrices, $I^i$, and the scalar product between the two Pauli spin vectors. We always use the operator form in calculations and not the matrix form. The off-diagonal coherent terms are responsible for the entanglement and preclude the product state. If those two coherent states are dropped, then the singlet state  becomes a product state:  
\begin{equation}\label{product2}
\begin{aligned}
	{{\rho }_{\psi _{12}^{-}}} &\xrightarrow[\text{off-diagonal}]{\text{drop}} {{\rho }^{12}_{\text{product}}}\\
	&={{\rho }_{1+}}{{\rho }_{2-}}+{{\rho }_{1-}}{{\rho }_{2+}}\\ &=\frac{1}{2}\left( {{I}^{1}}+\sigma _{Z}^{1} \right)\frac{1}{2}\left( {{I}^{2}}-\sigma _{Z}^{2} \right)+\frac{1}{2}\left( {{I}^{1}}-\sigma _{Z}^{1} \right)\frac{1}{2}\left( {{I}^{2}}+\sigma _{Z}^{2} \right) \\ 
	& =\frac{1}{4}\left( {{I}^{1}} {{I}^{2}}-\sigma _{Z}^{1} \sigma _{Z}^{2} \right)\text{ }	=\frac{1}{2}\left( \begin{matrix}
		0 & 0 & 0 & 0  \\
		0 & 1 & 0 & 0  \\
		0 & 0 & 1 & 0  \\
		0 & 0 & 0 & 0  \\
	\end{matrix} \right) \\  \\ 
\end{aligned}
\end{equation}
According to Bell's Theorem, \cite{BellBert}, product states cannot account for the violation of BI. However, rather than asserting that this means that the violation is due to non-locality, here we find it is due to the dropped coherent terms. The violation of BI is due entirely to the presence of a correlation due to coherence, thereby obviating non-locality. 

We define an EPR pair as the two particles that separate from a singlet state, and one moves to Alice and the other to Bob. We assume there is no entanglement between the two particles after separation.

To reflect the experimental configuration where the two filters are coplanar, we set $\phi_{ab} =0$,  $E\left( a,b \right)=-\cos {{\theta }_{ab}}\xrightarrow{\phi_{ab} =0}-\cos \left( {{\theta }_{a}}-{{\theta }_{b}} \right)$, where $E(a,b)$ represents the quantum correlation between EPR pairs. 
									
\subsubsection{The Local Singlet} \label{1sec101}
The expectation value for the correlation is defined by the quantum trace over the operators, \cite{neumann1}:
\begin{equation}\label{G3a}
	\left\langle AB \right\rangle =\text{Tr}\left( {{A}^{\dagger }}B{\rho } \right)
\end{equation} 
where  $\dagger$ is the operator adjoint. The correlation due to spin polarization between different filter settings is given by
\begin{equation}\label{G3}
	\begin{split}
		E\left( a,b \right)&=\underset{1,2}{\mathop{\text{Tr}}}\,\left( \sigma _{a}^{1}\sigma _{b}^{2}{{\rho }_{12}} \right)\\
		&=\mathbf{a}\cdot \left\langle {{\sigma }^{1}}{{\sigma }^{2}} \right\rangle \cdot \mathbf{b}
	\end{split}
\end{equation}
Remove the filter components $\mathbf{a}$ and $\mathbf{b}$ to focus on the expectation value only:
\begin{equation}\label{G4}
	\begin{split}
		\left\langle {{\sigma }^{1}}{{\sigma }^{2}} \right\rangle 
		&=-\frac{1}{2}\underset{1}{\mathop{\text{Tr}}}\,\left( {{\sigma 	}^{1}}{{\sigma }^{1}} \right){\mathop{\cdot \frac{1}{2}\underset{2}{\text{Tr}}}}\,\left( {{\sigma }^{2}}{{\sigma }^{2}} \right)\\
		&=-\underline{\underline{\mathbf{U}}}\cdot \underline{\underline{\mathbf{U}}}=-\underline{\underline{\mathbf{U}}}
	\end{split}
\end{equation}
using
\begin{equation}\label{U}
	\begin{aligned}
		\frac{1}{2}\text{Tr}\left( \sigma \sigma  \right)
		&=\frac{1}{2}\text{Tr}\left( {{\sigma }_{X}}{{\sigma }_{X}XX}+{{\sigma }_{Y}}{{\sigma }_{Y}YY}+{{\sigma }_{Z}}{{\sigma }_{Z}}ZZ \right) \\ 
		&=\frac{1}{2}\text{Tr}\left( I \right)\left( XX+YY+ZZ \right)\\
		&=\left( XX+YY+ZZ \right)\equiv \underline{\underline{\mathbf{U}}} \\ 
	\end{aligned}
\end{equation}
Here, $\underline{\underline{\mathbf{U}}}$ is the totally symmetric second-rank tensor that is the identity in 3D Cartesian space. Notice that the antisymmetric contributions in the product, $\sigma_i\sigma_j, i\ne j$, trace to zero and are~omitted.

We then find the usual result for the spin correlation:
\begin{equation}\label{G5}
	\begin{split}
		E\left( a,b \right)&=-\mathbf{a}\cdot \underline{\underline{\mathbf{U}}}\cdot \mathbf{b}=-\mathbf{a}\cdot \mathbf{b}\\
		&=-\cos \left( {{\theta }_{a}}-{{\theta }_{b}} \right)
	\end{split}
\end{equation}
which gives the maximum correlation from a singlet state. This calculation is performed before the singlet separates into an EPR pair.  The EPR paradox states that coincidence experiments, \cite{clauser,aspect,aspect2,zeilinger}, show that the entangled state appears to remain intact over spacetime after separation.  Bell's Theorem concludes, \cite{BellBert}, that only non-local connectivity between Alice and Bob can explain the violation of BI.  Here we show that it is due to the correlation between Alice and Bob's helicities.

From Equation (\ref{G4}), note the contraction between Alice and Bob. The only source of correlation between EPR pairs in this treatment is through this contraction. Alice and Bob must share the same body-fixed frame. This frame is determined as the EPR pair separates,~\cite{sanctuary3}.

\subsubsection{The Product State} \label{1sec102}
Dropping entanglement leaves the polarized states. Using Equation (\ref{product2}) gives the correlation from the product states, (pol):
\begin{equation}\label{G8}
	\begin{aligned}
		{{E}_{\text{pol}}}\left( a,b \right)=
		&\mathbf{a}\cdot \underset{1,2}{\mathop{\text{Tr}}}\,\left( {{\sigma }^{1}}{{\sigma }^{2}}{{\rho }^{12}_{\text{product}}} \right)\cdot \mathbf{b} \\ 
		& =-\mathbf{a}\cdot \frac{1}{2}\underset{1}{\mathop{\text{Tr}}}\,\left( {{\sigma }^{1}}\sigma _{Z}^{1} \right)\frac{1}{2}\underset{2}{\mathop{\text{Tr}}}\,\left( {{\sigma }^{2}}\sigma _{Z}^{2} \right)\cdot \mathbf{b} \\ 
		& =-\mathbf{a}\cdot ZZ\cdot \mathbf{b}=-\cos {{\theta }_{a}}\cos {{\theta }_{b}} \\ 
	\end{aligned}
\end{equation}
and this result can only satisfy, and not violate, BI. 

\subsubsection{The Source} \label{1sec103}
As the particles leave the source, the filters and their settings, $\mathbf{a}\text{ and }\mathbf{b}$, are far away, and we do not need the quantum trace:
\begin{equation}\label{G9}
	\begin{split}
		\left\langle {{\sigma }^{1}}{{\sigma }^{2}} \right\rangle \xrightarrow{\text{remove trace}}{{\sigma }^{1}}{{\sigma }^{1}}\cdot {{\sigma }^{2}}{{\sigma }^{2}}
	\end{split}
\end{equation}
and the operator describing a spin in free flight is the dyadic,  ${{\sigma }}{{\sigma }}$, see, e.g., Equation (\ref{G4}). This can be expressed using the well-known expression from Geometric Algebra, \cite{GA}:
\begin{equation}\label{G10}
	{\sigma}_i \sigma_j  ={{\delta}_{ij}}+{{\varepsilon }_{ijk}}i{{\sigma }_{k}}
\end{equation}
giving a bivector and the three components of the Levi-Civita tensor, $\underline{\underline{\underline{\varepsilon }}}$. The antisymmetric parts make no contribution to the spin polarization because they  trace out, see, e.g.,~\mbox{Equation (\ref{U})}.  To obtain those lost terms, we must find the correlation from the second term in Equation (\ref{G10}), which is a bivector, $i\sigma$, and renders spin complex, \cite{penrose1,penrose2}.

\subsubsection{The Helicity} \label{1sec104}	We define the geometric helicity as a second-rank operator by
\begin{equation}\label{G11}
	\underline{\underline{\mathbf{h}}}\equiv 	\underline{\underline{\underline{\varepsilon }}}\cdot i{{\sigma} }
\end{equation}
which is anti-Hermitian and rotates the plane perpendicular to the bivector, $i{{\sigma }}$. The helicity, $\underline{\underline{\mathbf{h}}}$, (we drop the subscript $g$ used earlier, $\underline{\underline{\mathbf{h}}}_{g}\to\underline{\underline{\mathbf{h}}}$)  is antisymmetric, odd to parity, indicating helicity spins left or right.
The correlation due to coherence (coh) in coincidence EPR experiments is defined and given by 
\begin{equation}\label{G12}
	E{{\left( a,b \right)}_{\text{coh}}}=\mathbf{a}\cdot \left\langle \underline{\underline{\mathbf{h}}}^1\cdot\underline{\underline{\mathbf{h}}}^2 \right\rangle \cdot \mathbf{b}=-\sin {{\theta }_{a}}\sin {{\theta }_{b}}
\end{equation}					
That is:
\begin{equation} 	
	\label{G13}
\begin{split}
		E{{\left( a,b \right)}_{\text{coh}}}
		&=\mathbf{a}\cdot \underset{12}{\mathop{\text{Tr}}}\,\left( \underline{\underline{{\mathbf{h}}}}^{1}{^\dagger}\cdot\underline{\underline{\mathbf{h}}}^{2}{{\rho }^{12}_\text{product}} \right)\cdot \mathbf{b}\\
		&= -\left(\mathbf{a}\cdot \underline{\underline{\underline{\varepsilon}}}\cdot\frac{1}{2}\underset{1}{\mathop{\text{Tr}}}\,\left( {{\sigma }^{1}}{{\sigma }_{Z}} \right)\right)\cdot \left(\mathbf{b}\cdot \underline{\underline{\underline{\varepsilon }} }\cdot\frac{1}{2}\underset{2}{\mathop{\text{Tr}}}\,\left({ {{\sigma }^{2}}{{\sigma }_{Z}}} \right)\right)\\
		&=- \left(\mathbf{a}\cdot \underline{\underline{\underline{\varepsilon }} }{\cdot\frac{1}{2}\underset{1}{\mathop{\text{Tr}}}\,\left( {{\sigma }^{1}}{{\sigma }^{1}} \right)\cdot Z}\right)\cdot \left(\mathbf{b}\cdot \underline{\underline{\underline{ }}}\underline{\underline{\underline{\varepsilon }} }\cdot\frac{1}{2}\underset{2}{\mathop{\text{Tr}}}\,\left(  {{\sigma }^{2}}{{\sigma }^{2}}\right) \cdot{Z}\right) \\
		&=-\left( \mathbf{a}\cdot \underline{\underline{\underline{\varepsilon }} }\cdot Z \right)\cdot \left( {\mathbf{b}\cdot \underline{\underline{\underline{\varepsilon }} }\cdot {Z}} \right)\\
		&=-\left( \mathbf{a}\times Z\right)\cdot \left( \mathbf{b}\times Z\right)\\
		&=-\sin {{\theta }_{a}}Y\cdot Y\sin {{\theta }_{b}}\\
		&=-\sin {{\theta }_{a}}\sin {{\theta }_{b}}
	\end{split}
\end{equation}
where the three vectors, $\left( \mathbf{a},\mathbf{b},Z \right)$, are coplanar for $\phi_{ab} = 0$. Assuming $\mathbf{a}$ and $\mathbf{b}$ lie in the $ZX$ plane, then $Y$ is the axis of linear momentum, which is the same for both Alice and Bob.

\subsubsection{The Conservation of Geometric Correlation} \label{1sec105}
The correlation between two spins in a local singlet is
\begin{equation}
	\label{G14}
	\begin{split}
		E\left( a,b \right)&=-\mathbf{a}\cdot \underline{\underline{\mathbf{U}}}\cdot \mathbf{b}\\&=-\cos \left( {{\theta }_{a}}-{{\theta }_{b}} \right)
	\end{split}
\end{equation}
The correlation from a separated EPR pair arises from both the polarization of the spins and their coherent helicity:
\begin{equation} \label{G15}
	\begin{split}
		E\left( a,b \right)&=-\mathbf{a}\cdot ZZ\cdot \mathbf{b}-\mathbf{a}\times Z\cdot \mathbf{b}\times Z\\
		&=-\cos {{\theta }_{a}}\cos {{\theta }_{b}}-\sin {{\theta }_{a}}\sin {{\theta }_{b}}\\
		&=-\cos \left( {{\theta }_{a}}-{{\theta }_{b}} \right)
	\end{split}
\end{equation}
The correlation is preserved between the entangled local singlet and the product state of a separated EPR pair, which can be expressed as being divided between the polarization and the coherence:
\begin{equation}\label{conservation}
 \overbrace{E\left( a,b \right)}^{\text{local}}=\overbrace{E{{\left( a,b \right)}_{\text{pol}}}+E{{\left( a,b \right)}_{\text{coh}}}}^{\text{separated}}\end{equation}
This articulates the Conservation of Geometric Correlation between an isotropic singlet and its EPR pair. The decomposition follows from finding the irreducible representations for the dyadic under some group.  Here, the decomposition is under the rotation group, \cite{bob}. The LHS of Equation (\ref{conservation}) expresses the correlation from an entangled singlet before separation. The RHS gives the correlation after separation. Since it is physically impossible to maintain entanglement beyond the local interaction between an EPR pair, upon separation, the two irreducible representations maintain the full entangled correlation without the need for non-locality.

There are several equivalent ways to express this, but its fundamental origin is the geometric product, Equation (\ref{G10}).  This result can also be generalized to $n$-tuples, \cite{multipole, bob}, and correlation is always conserved.

\section{Conclusions} \label{sec12}
Spin helicity, defined by a bivector, Equation (\ref{G11}), is an element of reality under quaternion symmetry, \cite{sanctuary1}. We suggest that spin is more fully defined by $\sigma \to \Sigma $, which includes both the Pauli vector and bivector and defines Q-spin as a non-hermitian operator:
\begin{equation}\label{complexspin}
\Sigma \equiv \sigma +{{\underline{\underline{\mathbf{h}}}}}
\end{equation}
Contracting with a unit vector gives a unit quaternion, $\mathbf{a}\cdot\Sigma$. Spin processes become a product of quaternions and hence a rotation. 

There is no bivector in the Dirac equation. However, it is shown elsewhere, \cite{sanctuary1}, that one is easily introduced. The main requirement for the Dirac field is that the gamma matrices must anticommute. Both $(\gamma^0,\gamma^1, \gamma^2, \gamma^3)$ and $(\gamma^0_s, \gamma^1_s, i\gamma^2_s, \gamma^3_s)$ satisfy this (the subscript ``$s$'' denotes spin spacetime, \cite{sanctuary1}). Dirac used the former and we use the latter.  After symmetry and parity arguments, we are led to two complementary spaces: a 2D polarization spacetime; a disc of angular momentum; and a space of quaternions, H, which spins that disc. 

In free flight, only helicity is manifest, \cite{sanctuary1}, but when encountering a probe field, the helicity ceases and the usual polarized Dirac spin with two states emerges. Helicity, $\underline{\underline{\mathbf{h}}}$, and angular momentum, $\sigma$, are complementary elements of reality. They epitomize the wave--particle duality of quantum theory.

Formulating spin this way, Equation (\ref{complexspin}), makes spin complex, \cite{penrose1,penrose2}, and gives it the structure of a spinning disc of polarization. It has four states: the usual two states of up and down when measured and two helicity states of L and R in free flight.  In contrast, the usual definition of helicity (as the projection of the spin vector onto the axis of linear momentum) is interpreted as giving the spin states of up and down.  That is, helicity in particle physics is not independent of spin, whereas here the helicity and spin polarization are distinct.

Bell's Inequalities show that no classical system can violate a bound. Bell's Theorem,~\cite{BellBert},  proves that to violate that bound, the classical variables must be non-local.  He then asserts that the observed violation by quantum systems must also be non-local. His mistake is assuming that classical and quantum variables are the same.  Whereas classical variables are real, commute, and form one convex set, quantum variables are complex, do not commute, and form two convex sets. As we discuss in greater detail in paper 4, \cite{sanctuary4}, the conclusion is stark:  not only is Bell's Theorem inapplicable, but little of Bell's work is relevant to quantum systems.

\textls[-5]{With the repudiation of Bell's Theorem, objections to the locality arguments of EPR,~\cite{EPR}}, no longer stand in the way of their conclusion: quantum mechanics gives an incomplete description of Nature. Rather, it is a theory of measurement restricted to our~spacetime. 

Non-locality is a concept beyond logic. Extending  entangled ``EPR'' channels over spacetime, \cite{Bennett}, to instantaneously connect distant partners, rather than them knowing their right from left, is something Occam would have little difficulty with.  The treatment here shows that the violation of BI is not evidence for non-locality \cite{BigBell}, but rather for local realism. 

\vspace{6pt} 





\funding{This research received no external funding.}

\dataavailability{No new data were created or analyzed in this study.}

\conflictsofinterest{The authors declare no conflicts of interest.}

\begin{adjustwidth}{-\extralength}{0cm}

\reftitle{References}
	 
\PublishersNote{}
\end{adjustwidth}
\end{document}